\documentclass[intlimits,twoside,a4paper]{article}

\usepackage{amsmath,amssymb}
\usepackage{graphicx}
\usepackage[T2A]{fontenc}
\usepackage[english]{babel}
\usepackage{multirow}

\usepackage[T2A]{fontenc}
\usepackage[cp1251]{inputenc}

\usepackage[eqsecnum]{cmpj2}

\issue{2014}{17}{1}{13501}
\doinumber{10.5488/CMP.17.13501}

\title{Application of Levin's transformations to virial series}
\author[C.C.F. Florindo, A.B.M.S. Bassi]{C.C.F. Florindo,
        A.B.M.S. Bassi\thanks{E-mail: bassi@iqm.unicamp.br}}
\address{Institute of Chemistry, University of Campinas --- {UNICAMP}, 13083--970  Campinas, Brazil
}

\authorcopyright{C.C.F. Florindo, A.B.M.S. Bassi, 2014}

\date{Received August 22, 2013, in final form October 15, 2013}

\begin{document}

\maketitle

\begin{abstract}
A new method of estimating high-order virial coefficients for fluids composed of equal three-dimensional rigid spheres is proposed. The predicted $B_{11}$ and $B_{12}$ values are in good agreement with reliable estimates previously reported. A new application of the Levin's transformations is developed, as well as a new way of using Levin's transformations is suggested. For the virial series of packing factor powers, this method estimates the $B_{13}$ value near 173.%
\pacs 51.30.+i, 64.30.+t
\end{abstract}

\section{Introduction}

The virial series is extremely important for obtaining accurate state equations, because it expands the compressibility factor of a fluid through a power series of an adequate variable. Using the packing factor, $\eta$,
\begin{equation}
Z = 1+ \sum_{i=2}^{\infty} B_{i}\eta^{(i-1)},  \label{101}
\end{equation}
where $Z$ is the compressibility factor and $B_{i}$ is the virial coefficient of order $i$. In the late nineteenth century, van der Waals \cite{waal01}, Boltzmann \cite{bolt02}, and van Laar  \cite{laar03}, analytically calculated the virial coefficients $B_{2}$, $B_{3}$, and $B_{4}$, of a gas formed by equal three-dimensional spherical rigid particles. So far, there are no analytical expressions to calculate the coefficients after $B_{4}$, even considering such simple particles. Thus, the best values for higher-order coefficients are obtained by numerical calculations of the Mayer's functions \cite{maye04}.

Using Mayer's functions, in 1953 the coefficient $B_{5}$ was obtained by Rosenbluth and Rosenbluth  \cite{rose05}. Subsequently, the coefficients $B_{6}$ and $B_{7}$ were calculated by Ree and Hoover  \cite{ree06}, and $B_{8}$ by van Rensburg  \cite{janse07}, and by Vlasov and You  \cite{lasov08}. $B_{9}$ was obtained by Lab\'{i}k and collaborators  \cite{labi09} in 2005, and $B_{10}$ by Clisby and McCoy in 2006  \cite{clis10}. Calculations of coefficients after $B_{10}$ were not performed, on account of the huge increase in the number of Mayer's diagrams and integrals to be analyzed. Thus, the coefficients subsequent to $B_{10}$ are estimated (see estimates in \cite{tian11}).

There are some methods reported in the literature for extrapolating the values of virial coefficients to orders higher than the tenth. Among them, stand out the Pad\'{e} approximants, the maximum entropy approximation, the density functional method, the series of continuous exponential, molecular dynamics and the differential approximation method. All these methods are considered to be very plausible. Nevertheless, unproved assumptions on the mathematical behavior of the series are imposed in their applications. Indeed, it is not even proved whether the 3D rigid sphere virial series does converge for all physically significant $\eta$ values, or does not. Then, to avoid such assumptions is a desirable aim.

Slow convergent or even divergent series frequently appear in problems involving the evaluation of integrals, solutions of differential equations, perturbation theory and others  \cite{bhow12}. Moreover, in many scientific problems, the series permits the computation of a small number of terms, which is not sufficient to obtain the required accuracy. In this context, the sequence transformations play an essential role, since they accelerate the convergence of the series without the need to compute higher-order terms  \cite{brezi13}.

In 1973, David Levin  \cite{levi14} introduced new sequence transformations, which improved the convergence of slowly convergent series. Moreover, this method is particularly suitable for the summation of strongly divergent series. According to Smith and Ford  \cite{smit15}, who compared the performance of several linear and nonlinear series transformations, the Levin-type ones are probably the most powerful and versatile convergence accelerators ever known. Baram and Luban  \cite{luba16} were the first to demonstrate the applicability of the Levin's transformations to the virial expansions of hard discs and rigid spheres  through estimates for $B_{7}$  \cite{luba17}. In recent years, many applications of Levin-type transformations have been reported in the literature, though the focus has been mainly in the field of quantum physics (see, for example, \cite{bhag18,bouf19,prun20,jetz21,king22,scott23}).

In this work, the Levin's transformations are used to estimate the $B_{11}$, $B_{12}$, and $B_{13}$, virial coefficients for gases composed of 3D equal rigid spheres, assuming the known values of the coefficients up to $B_{10}$. That is, a completely new methodology for estimating virial coefficients is proposed, and an unused way of using Levin's transformations is suggested. This work is organized into four more sections. In section~2, the Levin's transformations are briefly described, highlighting their mathematical structure. In section~3, the methodology used to estimate the virial coefficients is presented. In section~4, the obtained estimates are indicated and compared to those reported in the literature. Finally, in section~5 the results are commented.

\section{Levin's transformations}

In  this  section, the  mathematical background  of the  Levin's  acceleration  method is summarized. However, for a more detailed mathematical description of the method, references  \cite{weni24} and  \cite{home25} are suggested. The Levin's sequence transformations are applicable to the model sequence
\begin{equation}
\label{201}
s_{r}= s  + \omega_{r} \sum_{j=1}^{k} c_{j}/(r+\gamma)^{j-1},\qquad k, r \in \mathbb{N},
\end{equation}
where $k$ represents the order of the transformation, $\gamma$ is an arbitrary parameter which may not be a negative integer, $\omega_{r}$ is the remainder estimate and $s$ is the limit of the sequence when it converges, or the antilimit if it diverges. The convergence or divergence of the sequence depends on the behavior of $\omega_{r}$, for $ r\rightarrow\infty$.

\looseness=-1In equation~(\ref{201}) there are $k+1$ unknown quantities, that is, the limit or antilimit $s$ and the $k$ linear coefficients $c_{1},\dots ,c_{k}$. Thus, $k+1$ sequence elements $s_{r},\dots ,s_{r+k}$, and the corresponding remainder estimates $w_{r},\dots ,w_{r+k}$, are required for determining $s$. Evidently, imposing some kind of remainder estimate, as the three ones considered by Levin, most of the sequences do not follow this model sequence. However, in many cases for which $n\geqslant  N$, where $N\in \mathbb{N}$, the sequence can be considered of $k$th order, namely
\begin{equation}
\label{202}
s_{r}= s_{kn}  + \omega_{r} \sum_{j=1}^{k} c_{jn}/(r+\gamma)^{j-1}, \qquad n\leqslant  r\leqslant  n+k, \qquad n \geqslant  N,
\end{equation}
where $s= \lim_{n \to \infty} s_{kn}$. Thus, $s$ is both the limit of the sequences $\{s_{r}\}^{\infty}_{r=1}$ and $\{s_{kn}\}^{\infty}_{n=N}$,  whose convergence was accelerated relatively to $\{s_{r}\}^{\infty}_{r=1}$, or even created if $s$ is an antilimit of $\{s_{r}\}^{\infty}_{r=1}$.

According to Cramer's rule, the general Levin's transformation is
\begin{eqnarray}
\label{203}
\mathfrak{L}^{(n)}_{k} (\gamma,s_{r},\omega_{r})=\frac{\begin{vmatrix} s_{n}& \dots& s_{n+k}\\
\omega_{n}& \dots& \omega_{n+k}\\
\vdots & \ddots & \vdots & \\
\omega_{n}/(\gamma+n)^{k-1}& \dots& \omega_{n+k}/(\gamma + n+k)^{k-1}
\end{vmatrix}}{\begin{vmatrix} 1& \dots& 1\\
\omega_{n}& \dots& \omega_{n+k}\\
\vdots & \ddots & \vdots & \\
\omega_{n}/(\gamma+n)^{k-1}& \dots& \omega_{n+k}/(\gamma +n+k)^{k-1}
\end{vmatrix}}.
\end{eqnarray}
If the sequence elements satisfy equation~(\ref{201}), then the Levin's general sequence transformation is exact, i.e. $\mathfrak{L}^{(n)}_{k} (\gamma,s_{r},\omega_{r})=s$. But if they satisfy equation~(\ref{202}), then
\begin{equation}
\label{204}
\mathfrak{L}^{(n)}_{k} (\gamma,s_{r},\omega_{r})=s_{kn}\,.
\end{equation}
As a ratio of two determinants, the Levin's transformation is unsuitable for practical applications involving reliable evaluations of large order determinants. Therefore, alternative expressions are commonly employed. For example, considering the Vandermonde determinant, the equation~(\ref{203}) can be rewritten
\begin{eqnarray}
\label{205}
\mathfrak{L}^{(n)}_{k} (\gamma,s_{r},\omega_{r})= \frac{\sum_{j=0}^{k} (-1)^{j}\left(\begin{array}{c}k\\j\end{array}\right)\left(\frac{\gamma +n+j}{\gamma +n+k}\right)^{k-1}\frac{s_{(n+j)}}{\omega_{(n+j)}}}{\sum_{j=0}^{k} (-1)^{j}\left(\begin{array}{c}k\\j\end{array}\right)\left(\frac{\gamma +n+j}{\gamma +n+k}\right)^{k-1}\frac{1}{\omega_{(n+j)}}},\qquad k,r,n \in \mathbb{N}.
\end{eqnarray}

According to \cite{weni24}, the Levin's transformation should work very well for a given sequence $\left\{s_{r}\right\}$ if the sequence $\left\{\omega_{r}\right\}$ of the remainder estimates is chosen in such a way that $\omega_{r}$ is proportional to the dominant term of an asymptotic expansion of the remainder
\begin{eqnarray}
\label{206}
\varpi_{r}= s_{r}-s=\omega_{r}\left[c+O\left(r^{-1}\right)\right], \qquad r\rightarrow\infty.
\end{eqnarray}
However, $\omega_{r}$ is not determined by this asymptotic condition, so that it is possible to find a variety of sequences $\left\{\omega_{r}\right\}$ of the remainder estimates for a given sequence $\left\{s_{r}\right\}$. Thus, the practical problem that arises is how to find the sequence of the remainder estimates.

Based on purely heuristic arguments, Levin suggested  three kinds of the remainder estimates, $\omega_{r}$, for sequences of partial sums
\begin{eqnarray}
\label{207}
s_{r}=\sum_{i=1}^{r} a_{i}, \qquad r \in \mathbb{N}.
\end{eqnarray}
In the case of alternating partial sums, $s_{r}$, Levin suggested
\begin{eqnarray}
\label{208}
\omega_{r} =a_{r},\qquad r \in \mathbb{N}.
\end{eqnarray}
Substituting this relationship in equation~(\ref{205}), the Levin's \textit{t} transformation is obtained,
\begin{eqnarray}
\label{209}
t^{(n)}_{k} (\gamma,s_{r})= \frac{\sum_{j=0}^{k} (-1)^{j} \left(\begin{array}{c}k\\j\end{array}\right)\left(\frac{\gamma +n+j}{\gamma +n+k}\right)^{k-1}\frac{s_{(n+j)}}{a_{(n+j)}}}{\sum_{j=0}^{k} (-1)^{j} \left(\begin{array}{c}k\\j\end{array}\right)\left(\frac{\gamma +n+j}{\gamma +n+k}\right)^{k-1}\frac{1}{a_{(n+j)}}}\,.
\end{eqnarray}
In the case of a sequence of partial sums, $s_{r}$, satisfying a logarithmic convergence, i.e.,
\begin{eqnarray}
\label{210}
\lim_{r\to\infty} \frac{s_{r}+1-s}{s_{r}-s} = 1,
\end{eqnarray}
Levin suggested
\begin{eqnarray}
\label{211}
\omega_{r}=a_{r} (\gamma+r),\qquad r \in \mathbb{N},
\end{eqnarray}
which being substituted into equation~(\ref{205}) produces the Levin's \textit{u} transformation
\begin{eqnarray}
\label{212}
u^{(n)}_{k} (\gamma,s_{r})= \frac{\sum_{j=0}^{k} (-1)^{j} \left(\begin{array}{c}k\\j\end{array}\right)\frac{(\gamma +n+j)^{k-2}}{(\gamma +n+k)^{k-1}}\frac{s_{(n+j)}}{a_{(n+j)}}}{\sum_{j=0}^{k} (-1)^{j}\left(\begin{array}{c}k\\j\end{array}\right)\frac{(\gamma +n+j)^{k-2}}{(\gamma +n+k)^{k-1}}\frac{1}{a_{(n+j)}}}\,.
\end{eqnarray}
Finally, Levin also suggested
\begin{eqnarray}
\label{213}
\omega_{r}= \frac{a_{r}a_{r+1}}{a_{r}-a_{r+1}},\qquad r \in \mathbb{N},
\end{eqnarray}
which corresponds to the Levin's \textit{v} transformation
\begin{eqnarray}
\label{214}
v^{(n)}_{k} (\gamma,s_{r})= \frac{\sum_{j=0}^{k} (-1)^{j} \left(\begin{array}{c}k\\j\end{array}\right)\left(\frac{\gamma +n+j}{\gamma +n+k}\right)^{k-1}\frac{a_{(n+j)}-a_{(n+j+1)}}{a_{(n+j)}a_{(n+j+1)}}s_{n+j}}{\sum_{j=0}^{k} (-1)^{j} \left(\begin{array}{c}k\\j\end{array}\right)\left(\frac{\gamma +n+j}{\gamma +n+k}\right)^{k-1}\frac{a_{(n+j)}-a_{(n+j+1)}}{a_{(n+j)}a_{(n+j+1)}}}\,.
\end{eqnarray}
Other Levin-type transformations are reported in the literature (see  \cite{home25}), but only those originally proposed by Levin are listed above, and are used in this work.

\section{Methodology}

A Levin's transformation can be applied to a well-defined sequence to obtain a new sequence that presents a better convergence than the original one. Still, in this work, the sequence is not completely defined, and the supposition that some Levin's transformation can change the values of lower-order terms to the values of higher-order terms of the same sequence is tested for the virial series. Indeed, the virial series defined by equation~(\ref{101}) is a sequence of partial sums in accordance with equation~(\ref{207}), i.e.,
\begin{eqnarray}
\label{301}
s_{r}=\sum_{i=1}^{r} a_{i},\quad r \in \mathbb{N},\quad\mbox{and}\quad Z=\lim_{r\rightarrow\infty}s_{r}\,,
\end{eqnarray}
where $a_{i}= B_{i}\eta^{i-1}$ and $B_{1}=1$. In this case, $s_{r}$ is the value of the virial series truncated at the term proportional to $\eta^{r-1}$, whose coefficient is $B_{r}$. Then, it is supposed that the application of the Levin's transformation $\mathfrak{L}^{(n)}_{k}$ to $\{s_{r}\}^{\infty}_{r=1}$ produces, according to equation~(\ref{204}),
\begin{eqnarray}
\label{302}
\mathfrak{L}^{(n)}_{k} (\gamma,s_{r},\omega_{r})= s_{kn} = s_{n+k}\,,
\end{eqnarray}
i.e., the $n$th element of the sequence $\{s_{kn}\}^{\infty}_{n=N}$ is equal to the ($n+k$)th element of the sequence $\{s_{r}\}^{\infty}_{r=1}$, for all $n\geqslant  N$. Thus,
\begin{equation}
\label{303}
\mathfrak{L}^{(n)}_{k} (\gamma,s_{r},\omega_{r})=\sum_{i=1}^{n+k}B_{i}\eta^{i-1}.
\end{equation}

As already mentioned, the values of coefficients are precisely known up to $B_{10}$. But, for equations~(\ref{209}) and (\ref{212}), suppose that the first unknown term is
\begin{eqnarray}
\label{304}
a_{n+k}=B_{n+k}\eta^{n+k-1}\quad  \text{for all} \quad 2\leqslant  n+k\leqslant  10.
\end{eqnarray}
Then, using (\ref{303}) in equations~(\ref{209}) or (\ref{212}), \(a_{n+k}=B_{n+k}\eta^{n+k-1}\) can be found for all $n+k$ in \(2\leqslant  n+k\leqslant  10\). Analogously, for equation~(\ref{214}), suppose that the first unknown term is
\begin{eqnarray}
\label{305}
a_{n+k+1}=B_{n+k+1}\eta^{n+k}\quad \text{for all} \quad 2\leqslant  n+k\leqslant  9.
\end{eqnarray}
Using the equation~(\ref{303}) in ~(\ref{214}), \(a_{n+k+1}=B_{n+k+1}\eta^{n+k}\) can be found for all $n+k$ in \(2\leqslant  n+k\leqslant  9\).
In any case, for a given value of $\eta$, the corresponding virial coefficient can be calculated and compared with the values reported in the literature (table~\ref{tbl-1}). It is worthwhile noting that the choice of the values of  coefficients reported on table~1 has been made arbitrarily. Indeed, more recent references could be used, such as  \cite{wheat26}.
\begin{table}[ht]
\footnotesize
\caption{The virial coefficients $B_{i}$.}
\label{tbl-1}
\begin{center}
\renewcommand{\arraystretch}{0}
\begin{tabular}{|c|c|c|}
\hline
$i$ & \cite{labi09} &  \cite{clis10}\strut \\
\hline
\rule{0pt}{2pt}& & \\
\hline 1 & 1 & 1\strut \\
\hline 2 & 4 & 4\strut \\
\hline 3 & 10 & 10\strut \\
\hline 4 & 18.3647684 & 18.364768\strut \\
\hline 5 &  28.2245 $\pm$ 0.00010 &  28.2245 $\pm$  0.0003\strut \\
\hline 6 &  39.81550 $\pm$ 0.00036 &  39.81507 $\pm$ 0.00092\strut \\
\hline 7 & 53.3413 $\pm$ 0.0016 & 53.34426$\pm$ 0.00368\strut \\
\hline 8 &  68.540$\pm$ 0.010  &  68.538$\pm$ 0.018\strut \\
\hline 9 & 85.80$\pm$ 0.080 &85.813$\pm$ 0.085\strut \\
\hline 10 &  \ldots & 105.77$\pm$ 0.39\strut \\
\hline
\end{tabular}
\renewcommand{\arraystretch}{1}
\end{center}
\end{table}

In general, the estimation of a virial coefficient can be obtained from several representations of the same Levin's transformation, as shown in table~\ref{tbl-2}.
\begin{table}[ht]
\footnotesize
\caption{The representations which estimate the terms $a_{n+k}=B_{n+k}\eta^{n+k-1}$, $2\leqslant  n+k\leqslant  10$, by equations~(\ref{209}) (transformation $t$) or (\ref{212}) (transformation $u$), and $a_{n+k+1}=B_{n+k+1}\eta^{n+k}$, $2\leqslant  n+k\leqslant  9$, by equation~(\ref{214}) (transformation $v$, disregarding the last line in the table).}
\label{tbl-2}
\vspace{2ex}
\begin{center}
\renewcommand{\arraystretch}{0}
\begin{tabular}{|c|c|}
\hline
$i$ &  Levin's representations, $\mathfrak{L}^{n}_{k} (\gamma,s_{r},\omega_{r})$\strut \\
\hline
\rule{0pt}{2pt} & \\
\hline 2& $\mathfrak{L}^{1}_{1}$ \strut \\
\hline 3& $\mathfrak{L}^{1}_{2}$, $\mathfrak{L}^{2}_{1}$\strut \\
\hline 4& $\mathfrak{L}^{1}_{3}$, $\mathfrak{L}^{2}_{2}$, $\mathfrak{L}^{3}_{1}$\strut \\
\hline 5& $\mathfrak{L}^{1}_{4}$, $\mathfrak{L}^{2}_{3}$, $\mathfrak{L}^{3}_{2}$, $\mathfrak{L}^{4}_{1}$\strut \\
\hline 6& $\mathfrak{L}^{1}_{5}$, $\mathfrak{L}^{2}_{4}$, $\mathfrak{L}^{3}_{3}$, $\mathfrak{L}^{4}_{2}$, $\mathfrak{L}^{5}_{1}$\strut \\
\hline 7& $\mathfrak{L}^{1}_{6}$, $\mathfrak{L}^{2}_{5}$, $\mathfrak{L}^{3}_{4}$, $\mathfrak{L}^{4}_{3}$, $\mathfrak{L}^{5}_{2}$, $\mathfrak{L}^{6}_{1}$ \strut\\
\hline 8& $\mathfrak{L}^{1}_{7}$, $\mathfrak{L}^{2}_{6}$, $\mathfrak{L}^{3}_{5}$, $\mathfrak{L}^{4}_{4}$, $\mathfrak{L}^{5}_{3}$, $\mathfrak{L}^{6}_{2}$, $\mathfrak{L}^{7}_{1}$\strut\\
\hline 9& $\mathfrak{L}^{1}_{8}$, $\mathfrak{L}^{2}_{7}$, $\mathfrak{L}^{3}_{6}$, $\mathfrak{L}^{4}_{5}$, $\mathfrak{L}^{5}_{4}$, $\mathfrak{L}^{6}_{3}$, $\mathfrak{L}^{7}_{2}$, $\mathfrak{L}^{(8)}_{1}$\strut \\
\hline 10 & $\mathfrak{L}^{1}_{9}$, $\mathfrak{L}^{2}_{8}$, $\mathfrak{L}^{3}_{7}$, $\mathfrak{L}^{4}_{6}$, $\mathfrak{L}^{5}_{5}$, $\mathfrak{L}^{6}_{4}$, $\mathfrak{L}^{7}_{3}$, $\mathfrak{L}^{8}_{2}$, $\mathfrak{L}^{9}_{1}$\strut \\
\hline
\end{tabular}
\renewcommand{\arraystretch}{1}
\end{center}
\end{table}
The representations whose virial coefficients values do not deviate more than $1$\% from the corresponding values in table~\ref{tbl-1} are used to estimate higher-order coefficients. The representations do not provide good estimates for the coefficients of the order less than $B_{5}$, while for higher orders, acceptable values are found. This behavior stems from the lack in information supplied to the representations by the virial series truncated on the terms of the order smaller than the fourth, so that a minimum number of the known terms of the series is required.

The methodology is based on determining simple functions $\eta=f(i)$ ($i$ is the index of $B_{i}$) by using the optimal $\eta$ values which correspond to the best estimates of coefficients from $B_{5}$ to $B_{10}$. These functions are obtained both by interpolating the five or six optimal $\eta$ values themselves, and by interpolating their variations (optimal $\eta$ value for $B_{6}$ less optimal $\eta$ value for $B_{5}$, and so forth). The mathematical structures of such functions are determined by using the Mathematica computer program, version $8.0$. Once the functions are known, they are used to estimate $B_{11}$ and $B_{12}$.

\section{Estimates of the 11th, 12th and 13th virial coefficients}

In this section, the best representations and the corresponding estimates of coefficients are presented. For the \textit{t} and \textit{u} transformations, the $t^{n}_{3}$ and $u^{n}_{3}$ representations, respectively, provide good estimates of coefficients, while for the \textit{v} transformation, the best estimates are obtained through the $v^{n}_{2}$ representations. For the $t^{n}_{3}$ representations, the optimal $\eta$ for estimating the coefficients from $B_{5}$ to $B_{10}$ approximately lie between $0.20$ and $0.28$, while they are  approximately in the interval from $0.01$ to $0.08$ for the $u^{n}_{3}$ representations.

Using the $v^{n}_{2}$ representations, the optimal $\eta$ values are, approximately, in the interval from $0.40$ to $0.78$. This range is about five times broader than the other two ranges, yet a large $\eta$ variation for low $i$ values is not important, while the $\eta$ tendency to reduce its variation as the index $i$ increases is fundamental. Moreover, this range presents an upper bound about $5$\% greater than the physical one (the geometric maximum packing factor for rigid spheres is about $0.74$). Nonetheless, the $v^{n}_{2}$ representations are retained for this work, because this physical restriction is irrelevant for the present mathematical purpose. Moreover, packing factors above its physical upper bound, and even above $1.0$, are frequently considered in the literature.

\subsection{T-type representations}

Using the $t^{i-3}_{3}$ representations for $i=5,6,\dots ,12$, the virial coefficients $B_{i}$ are estimated. Thus, to predict the coefficients $B_{11}$ and $B_{12}$, the $t^{8}_{3}$ and $t^{9}_{3}$ representations are respectively used. The optimal $\eta$ values for the $t^{8}_{3}$ and $t^{9}_{3}$ representations are established by using four trial functions, which are obtained by interpolating the optimal $\eta$ values corresponding to the coefficients from $B_{5}$ to $B_{10}$. These optimal $\eta$ values are \(\eta=0.2493\) for $B_{5}$, \(\eta=0.2591\) for $B_{6}$, \(\eta=0.2703\) for $B_{7}$, \(\eta=0.2425\) for $B_{8}$, \(\eta=0.2172\) for $B_{9}$, and \(\eta=0.2053\) for $B_{10}$. Table~\ref{tbl-3} shows the values of the obtained virial coefficients, and their percentage deviations from the values reported in the literature.

According to table~\ref{tbl-3}, all $B_{12}$ estimates obtained from the interpolation of the optimal $\eta$ values themselves deviate more than 6\% from the reported values. Thus, the estimates  using the $\eta$ values from the functions of the optimal $\eta$ variations are the only ones providing virial coefficients close to those reported in the literature. Among the four functions, only the logarithmic and the straight-line functions provide deviations less than 3\% for both $B_{11}$ and $B_{12}$. Comparing these values with those obtained by Pad{\'e} approximants ($B_{11} = 128.6$ and $B_{12}=155$)  \cite{guerr27} one concludes that both methods lead to similar estimates. Therefore, this comparison, as well as the values reported in \cite{labi09} and \cite{clis10}, lead to the values of $B_{11}$ and $B_{12}$ obtained from the logarithmic and the straight-line functions of the optimal $\eta$ variations.

\begin{table}[htb]
\footnotesize
\caption{Values of the virial coefficients $B_{11}$ and $B_{12}$ estimated by the $t^{8}_{3}$ and $t^{9}_{3}$ representations, respectively. Percentage deviations from the values reported in the literature are also presented. $(\%)^{a}$ Percentage deviation from the $B_{11} = 129\pm 2$ and  $B_{12} = 155\pm 10$ values reported by \cite{labi09}. $(\%)^{b}$ Percentage deviation from the $B_{11} = 127.9$ and $B_{12}=152.7$ values reported by \cite{clis10}.}
\label{tbl-3}
\vspace{2ex}
\begin{center}
\renewcommand{\arraystretch}{0}
\begin{tabular}{|c||cccccc|cccccc|}
\hline
\raisebox{-1.7ex}[0pt][0pt]{Functions}
& \multicolumn{6}{c|}{$\eta$-variation}& \multicolumn{6}{c|}{$\eta$-absolute} \strut\\ \cline{2-13} &
$B_{11}$ & $(\%)^{ a}$ & $(\%)^{ b}$ & $B_{12}$ & $(\%)^{a}$ & $(\%)^{b}$ & $B_{11}$ & $(\%)^{ a}$ & $(\%)^{ b}$ & $B_{12}$ & $(\%)^{ a}$ & $(\%)^{ b}$ \strut\\ \hline
\rule{0pt}{2pt}&&&&&&&&&&&\\
\hline
Logarithmic &127.1 & 1.47 & 0.63 &156.5 & 0.97 & 2.49 & 133.8 & 3.72 & 4.61 & 170.2 & 9.81 & 11.5\strut \\
Exponential &131.3 & 1.78 & 2.66 & 164.9 & 6.39 & 7.99 & 131.1 & 1.63 & 2.50 & 165.5 & 6.77 & 8.38\strut \\
Straight-line & 126.2 & 2.17 & 1.33 &153.5 & 0.97 & 0.52 & 130.9 & 1.47 & 2.35 & 165.1 & 6.52 & 8.12\strut \\
Potency &126.8 & 1.71 & 0.86 &147.4 & 4.90 & 3.47 & 133.6 & 3.57 & 4.46 & 169.9 & 9.61 & 11.3\strut \\
\hline
\end{tabular}
\renewcommand{\arraystretch}{1}
\end{center}
\end{table}

Thus, the logarithmic and straight line functions of the optimal $\eta$ variations are also considered to estimate the value of $B_{13}$.  Using  the $t^{10}_{3}$ representation, the values $183.68$ and $175.45$ are respectively found. The value from the logarithmic function is very close to those estimated in \cite{clis10} ($181.19$) and  \cite{tian11} ($180.82$), whereas the value obtained from the straight-line function is between those estimated in \cite{coy28} ($177.40$) and  \cite{santo29} ($171.28$). Therefore, this comparison confirms that the logarithmic and straight-line functions can be used to find estimates of the optimal $\eta$ for high order coefficients.
\\
\subsection{U-type representations}

Using the $u^{i-3}_{3}$ representations for $i=5,6,\dots ,12$, the virial coefficients $B_{i}$ are also estimated. Thus, to predict the coefficients $B_{11}$ and $B_{12}$ the $u^{8}_{3}$ and $u^{9}_{3}$ representations are respectively used. The optimal $\eta$ values are $0.01714$ for $B_{6}$, $0.06435$ for $B_{7}$, $0.07235$ for $B_{8}$, $0.07209$ for $B_{9}$, and $0.07819$ for $B_{10}$ (the $1$\% minimal deviation is not attained for $B_{5}$). Table~\ref{tbl-4} presents the values of the obtained virial coefficients.

\begin{table}[htb]
\footnotesize
\caption{Values of the virial coefficients $B_{11}$ and $B_{12}$ estimated by the $u^{8}_{3}$ and $u^{9}_{3}$ representations, respectively. Percentage deviations from the values reported in the literature are also presented.}
\label{tbl-4}
\vspace{2ex}
\begin{center}
\renewcommand{\arraystretch}{0}
\begin{tabular}{|c||cccccc|cccccc|}
\hline
\raisebox{-1.7ex}[0pt][0pt]{Functions}
& \multicolumn{6}{c|}{$\eta$-variation}& \multicolumn{6}{c|}{$\eta$-absolute} \strut\\ \cline{2-13} &
$B_{11}$ & $(\%)^{ a}$ & $(\%)^{ b}$ & $B_{12}$ & $(\%)^{a}$ & $(\%)^{b}$ & $B_{11}$ & $(\%)^{ a}$ & $(\%)^{ b}$ & $B_{12}$ & $(\%)^{ a}$ & $(\%)^{ b}$ \strut\\ \hline
\rule{0pt}{2pt}&&&&&&&&&&&\\
\hline
\\
Logarithmic & 127.3 & 1.32 & 0.47 & 150.3 & 3.03 & 1.57 & 131.0 & 1.55 & 2.42 & 162.8 & 5.03 & 6.61\strut \\
Exponential & 128.9 & 0.08 & 0.78 & 156.6 & 1.03 & 2.55  & 132.9 & 3.02 & 3.91 & 170.9 & 10.3 & 11.9\strut \\
Straight-line & 126.2 & 2.17 & 1.33 & 221.5 & 42.9 & 45.1 & 132.4 & 2.64 & 3.52 & 167.9 & 8.32 & 9.95\strut \\
Potency & 129.1 & 0.08 & 0.94 & 156.9 & 1.23 & 2.75  & 131.5 & 1.94 & 2.82 & 164.5 & 6.13 & 7.73\strut \\
\hline
\end{tabular}
\renewcommand{\arraystretch}{1}
\end{center}
\end{table}

Table~\ref{tbl-4} shows that the $\eta$ values from the logarithmic, exponential and potency functions obtained by interpolating the optimal $\eta$ variations provide good estimates. In the case of a straight-line function, one can also note a good estimate, but only for $B_{11}$.  Considering the $\eta$ values obtained from functions of the optimal $\eta$ values themselves, only the logarithmic function provides a $B_{12}$ value which deviates less than $6$\% from a reported value. This function also provides the best estimate for $B_{11}$. However, imposing the smaller percentage deviations as a criterion, the logarithmic, exponential and potency functions obtained by interpolating the optimal $\eta$ variations are selected to estimate high-order coefficients.

The $B_{13}$ values estimated by using the exponential and potency functions of the optimal $\eta$ variations are 190.03 and 190.80, respectively. It is impossible to obtain an estimate of $B_{13}$ from the logarithmic function, because the $u^{10}_{3}$ representation does not support the supplied optimal $\eta$ value. The estimated values are above those obtained by \cite{clis10,tian11,coy28,santo29}. However, they are close to the values estimated in \cite{kola30} (190.82) and  \cite{labi31} (185$\pm$10).
\\
\subsection{V-type representations}

In the last test, the  $v^{8}_{2}$ and $v^{9}_{2}$ representations are used to estimate the coefficients $B_{11}$ and $B_{12}$. The optimal $\eta$ values are $0.4010$ for $B_{5}$, $0.5445$ for $B_{6}$, $0.6507$ for $B_{7}$, $0.7089$ for $B_{8}$, $0.7464$ for $B_{9}$, and $0.7783$ for $B_{10}$. Table~\ref{tbl-5} presents the estimated coefficients $B_{11}$ and $B_{12}$, as well as their deviations from previously reported values. This table clearly indicates that the $\eta$ values corresponding to the potency and exponential functions of the optimal $\eta$ variations, and the logarithmic interpolation of the optimal $\eta$ values themselves, provide the best estimates for the coefficients $B_{11}$ and $B_{12}$. Using the $v^{10}_{2}$ representation and the potency function of the variations, which shows the lowest deviations, the coefficient $B_{13}$ is estimated. The value obtained is $172.65$. This estimated value is lower than those presented in  \cite{clis10,tian11,kola30,labi31}, but it is placed between those in  \cite{coy28} and  \cite{santo29}.

\begin{table}[htb]
\footnotesize
\caption{Values of the virial coefficients $B_{11}$ and $B_{12}$ estimated by the $v^{8}_{2}$ and $v^{9}_{2}$ representations, respectively. Percentage deviations from the values reported in the literature are also presented.}
\label{tbl-5}
\vspace{2ex}
\begin{center}
\renewcommand{\arraystretch}{0}
\begin{tabular}{|c||cccccc|cccccc|}
\hline
\raisebox{-1.7ex}[0pt][0pt]{Functions}
& \multicolumn{6}{c|}{$\eta$-variation}& \multicolumn{6}{c|}{$\eta$-absolute} \strut\\ \cline{2-13} &
$B_{11}$ & $(\%)^{ a}$ & $(\%)^{ b}$ & $B_{12}$ & $(\%)^{a}$ & $(\%)^{b}$ & $B_{11}$ & $(\%)^{ a}$ & $(\%)^{ b}$ & $B_{12}$ & $(\%)^{ a}$ & $(\%)^{ b}$ \strut\\ \hline
\rule{0pt}{2pt}&&&&&&&&&&&\\
\hline
Logarithmic & 131.5 & 1.94 & 2.81 & 166.5 & 7.42 & 9.04 & 127.8 & 0.93 & 0.08 & 149.1 & 3.81 & 2.36\strut \\
Exponential & 130.8 & 1.39 & 2.27 & 162.2 & 4.52 & 6.09 &  116.6 & 10.1 & 8.84 & 106.8 & 31.1 & 30.1\strut \\
Straight-line & 135.1 & 4.73 & 5.63 & 187.6 & 21.1& 22.9  & 119.2 & 7.60 & 6.80 & 117.1 & 24.5 & 23.3\strut \\
Potency & 128.9 & 0.08 & 0.78 & 152.5 & 1.61 & 0.13 &124.9 & 3.18  & 2.35 & 137.7 & 11.2 & 9.82 \strut \\
\hline
\end{tabular}
\renewcommand{\arraystretch}{1}
\end{center}
\end{table}

\section{Conclusions}

Only representations with $k=2$ for the Levin's \textit{v} transformation, and $k=3$ for the \textit{t} and \textit{u} Levin's transformations, are acceptable. This interesting result guided the choice of the representations used to estimate the high order virial coefficients. Moreover, the estimates also depend on the dimensionless $\eta$ value. This is an expected dependence, because the Levin's convergence accelerators modify the terms from the series, not just the coefficients included within these terms.

The $B_{11}$ and $B_{12}$ values have been confirmed in the literature, by using distinctive methodologies, which imply different assumptions on the mathematical behavior of the virial series. Thus, such values are reliable. Meanwhile, the values reported in the literature are in accordance with some representations of the Levin's transformations, highlighting these transformations usefulness in the prediction of virial coefficients.

Should a Levin's transformation be able to change the values of lower order terms to the values of higher-order terms of the considered virial series, then it is expected that: (i) such ability is enhanced for high order terms of the series, which are favored by high $\eta$ values, and (ii) the $\eta$ value variation caused by substituting $i+1$ for $i$ decreases as $i$ increases ($\eta$ tends to some unique value for high-order coefficients). Accordingly, note that, in table~\ref{tbl-3} to~\ref{tbl-5}, the functions are not selected to achieve the best fit to the $B_{5}$ to $B_{10}$ values in table~\ref{tbl-1} (for instance, functions with more than two parameters are not used), but to test the asymptotic behavior of the functions. An interesting result is that all the functions selected in section~4 by comparing the obtained values to previously reported ones, except the straight line function, are asymptotic to the $i$ axis, that is, they satisfy the above condition (ii).

The $\eta$ values corresponding to the $u$ transformation refer to the gaseous state, whose description is accurate enough by using only the low order terms of the series, and the $\eta$ values corresponding to the $t$ transformation concern the liquid state, whose description is accurate enough by using the low and medium order terms of the series. However, the $\eta$ values corresponding to the $v$ transformation refer to the overcooled liquid and vitreous states, whose accurate descriptions also demand the high order terms of the series. Note that the $v$ transformation is the only one producing good results not exclusively from the interpolation of the $\eta$ values variations, but also from the interpolation of the optimal $\eta$ values themselves, confirming the above condition (i). Thus, the $v$ transformation, which is the least specific one among those originally presented by Levin, is preferable. As a consequence of this choice, the $B_{13}$ value near $173$ is proposed in this work. Note that the 13 terms long, virial series for rigid spheres developed in this work should be useful in describing the repulsive pressure of overcooled liquids and vitreous transitions. However, this series will not reproduce crystallization, which involves drastic changes in entropy and volume.

\section*{Acknowledgement}
Financial support by the Brazilian federal government agency --- CAPES --- and the University of Campinas~--- UNICAMP~--- are acknowledged.

{\small \topsep 0.6ex

}

\ukrainianpart

\title{Застосування перетворень Левіна до віріальних рядів}

\author{С.С. Фльоріндо, А.Б.М.С. Бассі}

\address{Інститут хімії, університет м. Кампінас ---
{UNICAMP}, 13083--970  Кампінас, Бразилія}

\makeukrtitle

\begin{abstract}
\tolerance=3000%
Запропоновано новий метод оцінювання віріальних коефіцієнтів
високого порядку для плинів, що складаються з однакових тривимірних
жорстких сфер.  Передбачені значення для $B_{11}$ і $B_{12}$ добре
узгоджуються з надійними оцінками, отриманими раніше. Розвинуто нове
застосування перетворень Левіна, а також запропоновано новий спосіб
використання перетворень Левіна. Для віріальних рядів за степенями
упаковання сфер цей метод дає оцінку для значення  $B_{13}$ близько 173.%

\keywords перетворення Левіна, віріальні ряди, жорсткі сфери

\end{abstract}


\begin{thebibliography}{99}
\bibitem{waal01} van der Waals~J.D., Proc. K. Ned. Akad. Wet., 1899, \textbf{1}, 138.

\bibitem{bolt02} Boltzmann~L., Versl. Gewone Vergad. Afd. Natuurkd., K. Ned.
Akad. Wet., 1899, \textbf{7}, 484.

\bibitem{laar03} van Laar~J.J., Proc. K. Ned. Akad. Wet., 1899, \textbf{1}, 273.

\bibitem{maye04} Mayer~J.E., Mayer~M.G., Statistical Mechanics, John Wiley, New York, 1940.

\bibitem{rose05} Rosenbluth~M.N., Rosenbluth~A.W., J. Chem. Phys., 1955, \textbf{23}, 356; \doi{10.1063/1.1741967}.

\bibitem{ree06} Ree~F.H., Hoover~W.G., J. Chem. Phys., 1967, \textbf{46}, 4181; \doi{10.1063/1.1840521}

\bibitem{janse07} van Rensburg~E.J., J. Phys. A: Math. Gen., 1993, \textbf{26}, 4805; \doi{10.1088/0305-4470/26/19/014}.

\bibitem{lasov08} Vlasov~A.Y.,  You~X.M., Masters~A.J., Mol. Phys., 2002, \textbf{100}, 3313; \doi{10.1080/00268970210153754}.

\bibitem{labi09} Lab{\'i}k~S., Kolafa~J., Malijevsk{\'y}~A., Phys. Rev. E, 2005, \textbf{71}, 021105; \doi{10.1103/PhysRevE.71.021105}.

\bibitem{clis10} Clisby~N., McCoy~B.M., J. Stat. Phys., 2006, \textbf{122}, 15; \doi{10.1007/s10955-005-8080-0}.

\bibitem{tian11} Tian~J.X., Jiang~H., Gui~Y.X., Mulerlo~A.,  Phys. Chem. Chem. Phys., 2009, \textbf{11}, 11213; \doi{10.1039/b915002a}.

\bibitem{bhow12} Roy~D., Bhattacharya~R., Bhowmick~S.,  Comp. Phys. Commun., 1979, \textbf{93}, 159;  \doi{/10.1016/0010-4655(95)00106-9}.

\bibitem{brezi13} Brezinski~C., Zaglia~M.R.,  Extrapolation Methods: Theory and Practice (Studies in Computational Mathematics), North  Holland, Amsterdam, 1991.

\bibitem{levi14} Levin~D., Int. J. Comput. Math., 1973, \textbf{3}, 371; \doi{10.1080/00207167308803075}.

\bibitem{smit15} Smith~D.A., Ford~W.F., Math. Comput., 1982, \textbf{38}, 481; \doi{10.2307/2007284}.

\bibitem{luba16} Baram~A., Luban~M.,  J. Phys. C: Solid State Phys., 1979, \textbf{12}, L659; \doi{10.1088/0022-3719/12/17/005}.

\bibitem{luba17} Erpenbeck~J.F., Luban~M.,  Phys. Rev. A, 1985, \textbf{32}, 2920;  \doi{10.1103/PhysRevA.32.2920}.

\bibitem{bhag18} Bhagat~V., Bhattacharya~R., Roy~D.,  Comput. Phys. Commun., 2003, \textbf{155}, 7;  \doi{10.1016/S0010-4655(03)00294-7}.

\bibitem{bouf19} Bouferguene~A., Fares~M., Phys. Rev. E, 1994, \textbf{49}, 3462; \doi{10.1103/PhysRevE.49.3462}.

\bibitem{prun20} de Prunel{\'e}~E., Int. J. Quantum Chem., 1997, \textbf{63}, 1079;  \doi{10.1002/(SICI)1097-461X(1997)63:6<1079::AID-QUA2>3.0.CO;2-U}.

\bibitem{jetz21} Jetzke~S., Broad~J.T., Int. J. Mod. Phys. C, 1991, \textbf{2}, 377; \doi{10.1142/S0129183191000524}.

\bibitem{king22} King~F.W., Int. J. Quantum Chem., 1999, \textbf{72}, 93;  \doi{ 10.1002/(SICI)1097-461X(1999)72:2<93::AID-QUA2>3.0.CO;2-#}.

\bibitem{scott23} Scott~T.C., Aubert-Fr{\'e}con~M., Andrae D., Appl. Algebr. Eng. Commun. Comput., 2002, \textbf{13}, 233;  \doi{10.1007/s002000200100}.

\bibitem{weni24} Weniger~E.J., Comput. Phys. Rep., 1989, \textbf{10}, 189; \doi{10.1016/0167-7977(89)90011-7}.

\bibitem{home25} Homeier~H.H.H., J. Comput. Appl. Math., 2000, \textbf{122}, 81; \doi{10.1016/S0377-0427(00)00359-9}.

\bibitem{wheat26} Wheatley~R.J., Phys. Rev. Lett., 2013, \textbf{110}, 200601; \doi{10.1103/PhysRevLett.110.200601}.

\bibitem{guerr27} Guerrero~A.O., Bassi~A.B.M.S., J. Chem. Phys., 2008, \textbf{129}, 044509; \doi{10.1063/1.2958914}.

\bibitem{coy28} Clisby~N., McCoy~B. M., Pramana-J. Phys., 2005, \textbf{64}, 775; \doi{10.1007/BF02704582}.

\bibitem{santo29} Santos~A., L{\'o}pez de Haro~M., J. Chem. Phys., 2009, \textbf{130}, 214104; \doi{10.1063/1.3147723}.

\bibitem{kola30} Kolafa~J., Lab{\'i}k~S.,  Malijevsky~A.,  Phys. Chem. Chem. Phys., 2004, \textbf{6}, 2335; \doi{10.1039/b402792b}.

\bibitem{labi31} On\v{c}{\'a}k~M., Malijevsk{\'y}~A., Kolafa~J., Lab{\'i}k~S., Condens. Matter Phys., 2012, \textbf{15}, 23004;  \doi{10.5488/CMP.15.23004}.

\end{thebibliography}
\end{document}